# Spin waves and orbital contribution to ferromagnetism in a topological metal

Wenliang Zhang[1], Teguh Citra Asmara[1, 2], Yi Tseng[1], Junbo Li[3], Yimin Xiong[4, 5], Yuan Wei[1], Tianlun Yu[1], Carlos William Galdino[1], Zhijia Zhang[1], Kurt Kummer[6], Vladimir N. Strocov[1], Y. Soh[1*], Thorsten Schmitt[1†], and Gabriel Aeppli[1, 7, 8, 9]

*[1]Paul Scherrer Institut, CH-5232 Villigen PSI, Switzerland*

*[2]European X-Ray Free-Electron Laser Facility GmbH, Schenefeld, Germany*

*[3]Anhui Province Key Laboratory of Condensed Matter Physics at Extreme Conditions, High Magnetic Field Laboratory, Chinese Academy of Sciences, Hefei 230031, China*

*[4]Department of Physics, School of Physics and Optoelectronics Engineering, Anhui University, Hefei 230601, China*

*[5]Hefei National Laboratory, Hefei 230028, China*

*[6]European Synchrotron Radiation Facility, 71 Avenue des Martyrs, Grenoble F-38043, France*

*[7] Department of Physics, ETH Zurich, CH-8093 Zurich, Switzerland.*

*[8]Quantum Center, ETH Zurich, CH-8093 Zurich, Switzerland.*

*[9]Institut de Physique, EPF Lausanne, CH-1015 Lausanne, Switzerland.*

To whom correspondence should be addressed: *yona.soh@psi.ch; †thorsten.schmitt@psi.ch

**Abstract**
Special arrangements of atoms with more than one atom per unit cell, including honeycomb or kagome (woven bamboo mat) lattices, can host propagating excitations with non-trivial topology as defined by their evolution along closed paths in momentum space. Excitations on such lattices can also be momentum-independent, meaning that they are localized notwithstanding strong hopping of the underlying disturbances between neighbouring sites. The associated flat bands are interesting because the interactions between the heavy quasiparticles inhabiting them will become much more important than for strong dispersion, resulting in novel quantum solid and liquid states. Different stackings of two-dimensional lattices, for example twisted graphene bilayers, provide routes to further engineer topology and many-body effects. Here, we report the discovery, using circularly polarized x-rays for the unambiguous isolation of magnetic signals, of a nearly flat spin wave band and large (compared to elemental iron) orbital moment for the metallic ferromagnet $Fe_3Sn_2$, built from compact AB-stacked kagome bilayers and which has a topologically non-trivial electronic band structure controllable by modest external magnetic fields. As a function of out-of-plane momentum, the nearly flat optical mode and the global rotation symmetry-restoring acoustic mode are out of phase, consistent with a bilayer exchange coupling that is larger than the already large in-plane couplings. The defining units of this topological metal are therefore a triangular lattice of octahedral iron clusters rather than weakly coupled kagome planes. The spin waves are strongly damped when compared to elemental iron, opening the topic of interactions of topological bosons (spin waves) and fermions (electrons) with the very specific target of explaining boson lifetimes.

## Introduction

Technology has long relied on transition metal-based ferromagnets, with their high Curie temperatures and many other convenient properties, such as tunable hardness, which allows a range of applications from motors to electrical transformer cores. The basic understanding of the magnetism of such compounds needed to await the development of the quantum theory of metals in the mid-twentieth century, but there continue to be surprises concerning fundamental properties such as the voltages developed transverse to electrical currents. An example of an intermetallic compound consisting of two very common elements – iron and tin – which challenges even the contemporary quantum theory of solids is $Fe_3Sn_2$.

$Fe_3Sn_2$ is a ferromagnet with a high Curie point $T_C \approx 640$ K[1–3] and consists of kagome bilayers stacked along $\mathbf{c}$ with the crystal structure belonging to the space group R-3m(1)[4]. The kagome layers are composed of two different sets of equilateral triangles with different Fe-Fe distances as indicated by magenta and blue bonds in Figure 1a[4], and are stacked with an offset along the (1, -1) in-plane lattice direction. Two key questions about $Fe_3Sn_2$ follow from the initial impression, based on the high temperature metallic ferromagnetism itself and the substantial Fe content, that the material is simply a diluted version of elemental iron.

The first concerns the orbital angular momentum $\mathbf{L}$ in $Fe_3Sn_2$ that is almost entirely quenched for Fe but has not been directly measured for $Fe_3Sn_2$ notwithstanding numerous thermodynamic and electrical properties which can only begin to be understood if $\mathbf{L}$ and the spin-orbit coupling are taken into account. Most prominent are a large anomalous Hall effect[5,6] and a transition around 120 K where the preferred magnetization direction rotates from $\mathbf{c}$ towards the kagome planes on cooling[1,2,7]. The latter has been a topic since 1970, with recent work unequivocally confirming that it is of first order and could be explained by the crossing of electronic free energies for the different magnetization directions[8,9]. Theory including the spin-orbit interaction as a key ingredient predicts that new phases, including states with fractionally charged quasiparticles, could emerge due to strong interactions for electrons occupying flat bands[10–12]. Recently, indications of fractionalized charge at zero magnetic field have been reported for $Fe_3Sn_2$[13]. Other discoveries center on the very large number of topologically non-trivial band crossings of Weyl character, within 10 meV of the Fermi level[14,15]; for iron there are a handful of Weyl nodes but they are further away from the Fermi level[16]. Density functional calculations show that the disposition of Weyl nodes near the Fermi level for $Fe_3Sn_2$ responds strongly to rotations of the magnetization, which accounts for the manipulation of electronic properties via modest external fields[14,17–19].

The second question is whether there is any physical property of three-dimensional $Fe_3Sn_2$ approaching that of an ideal kagome ferromagnet with short-range interactions, including Dirac crossings and perfectly flat bands among the spin waves, which, analogous to the electronic bands, can become topologically non-trivial with additional anisotropic or antisymmetric interactions. Such topological magnons are indeed observed in a quasi 2D kagome ferromagnetic insulator – the metal-organic compound Cu[1,3-benzenedicarboxylate (bdc)][20]. With different stackings of the kagome planes, the bands and topology are further modified, which can lead to new phenomena[21–25]. On the experimental front, topological Dirac magnons but no flat bands have been reported for a room-temperature magnet $YMn_6Sn_6$[26] consisting of ferromagnetic kagome double layers stacked with a simple vertical shift (AA-stacking).

Fig. 1a illustrates that the AB bilayers in $Fe_3Sn_2$ are quite compact with the short 2.584 Å inter-planar Fe-Fe bonds (yellow) nearly the same as the shorter 2.582 Å in-plane bonds (magenta) characterizing the breathing kagome planes with the alternating longer 2.732 Å bonds (blue)[3], yielding nearly perfect Fe octahedra. The bilayers are sandwiched between honeycomb Sn layers with a much longer inter-bilayer than intra-bilayer distance, leading to possible quasi-2D behavior of the spin degrees of freedom (DOF) on the Fe atoms. The bilayer structure suggests that the most relevant magnetic interactions in $Fe_3Sn_2$ are the in-plane nearest-neighbour interaction $J_{1a}$ ($J_{1b}$) in the small (large) triangles and the bilayer interaction $J_{bi}$. Neglecting the difference between $J_{1a}$ and $J_{1b}$, $i.e.$, setting $J_{1a} = J_{1b} = J_1$, the spin Hamiltonian of the bilayer structure can be expressed as:

$$\mathcal{H} = \sum_{n=1,2} \sum_{\langle i,j \rangle} J_1 \, \mathbf{S}_i^n \cdot \mathbf{S}_j^n + \sum_{\langle i,j \rangle_{1,2}} J_{bi} \, \mathbf{S}_i^1 \cdot \mathbf{S}_j^2 \qquad (1)$$

In (1) n = 1, 2 is a layer index, $\langle i,j \rangle$ labels nearest-neighbour pairs within a layer, and $\langle i,j \rangle_{1,2}$ labels nearest-neighbour pairs between the two layers. The spin wave structure is then determined by the ratio $J_{bi}/J_1$. Fig. 1c-f show the dispersion and spin-spin correlations simulated by the spinW package[27] for S = 1 with three characteristic values of $J_{bi}/J_1$, namely, the weak bilayer coupling limit (c) which reduces to the single kagome plane for $J_{bi}$=0, the intermediate bilayer coupling regime (d) for $J_{bi} \sim J_1$, and the strong bilayer coupling limit (e) which reduces to independent octahedral molecules of Fe arranged on a triangular (not kagome) lattice for $J_{bi}/J_1 = \infty$. The bilayer coupling will lift the parity degeneracy, splitting the spin wave bands into even and odd modes with opposite out-of-plane momentum dependence[28,29] as shown in Fig. 1g. On increasing the bilayer coupling, the spin waves show varied band crossings as a function of both momentum and $J_{bi}/J_1$ as the split OP1 band evolves from the lowest optical mode to the highest in the strong bilayer coupling limit. Fig. 1f shows the 2D dispersion for $J_{bi}/J_1 = 1.5$, which has many crossing points in the optical bands. With additional anisotropic or antisymmetric interactions, gaps can be opened at these points.

To summarize, notwithstanding many experimental and theoretical papers dealing with $Fe_3Sn_2$, there are no clearly resolved spectroscopic data confirming an impact of the kagome lattice geometry on either charge or spin degrees of freedom. In addition, the relative orbital and spin contributions to the magnetism, as well as the underlying spin Hamiltonian, are unknown. We have therefore used X-ray magnetic circular dichroism (MCD) to establish the orbital component of the magnetic moment, and to perform resonant inelastic scattering (RIXS) to determine the magnetic Hamiltonian. The MCD X-ray absorption spectroscopy (XAS) data reveal a much larger orbital contribution to the ferromagnetism than for elemental iron. Furthermore, we identify with MCD RIXS[30] both conventional acoustic spin waves, with stiffness similar to that found at very low energy transfers (< 2 meV) from neutron scattering from powders[31], as well as an optical mode with at most a weak dispersion. The two spin wave modes display out-of-phase intensities as a function of out-of-plane momentum, consistent with even and odd modes (Fig. 1g) induced by the bilayer interaction $J_{bi}$. The fitting to the linear spin wave theory suggests $J_{bi} \approx 1.5 \, J_1$, directly implying that $Fe_3Sn_2$ is far from the weakly-coupled kagome bilayer limit. Furthermore, the damping of both acoustic and optical modes is considerable at all studied momenta, indicating unusually strong interactions with electron-hole pairs.

**Experiment and results**

Fig. 1b shows the experimental geometry for XAS and RIXS measurements, where the circular polarization of the incident beam is specified while the outgoing polarizations are not resolved. To observe MCD, the magnetic domains in the sample need to be aligned, for which purpose we apply a magnetic field (~ 0.13 T) along the $(H, 0, 0)$ direction lying in the kagome planes. Fig. 2 summarizes the MCD XAS results for the $Fe_3Sn_2$ sample. The data are collected by the total electron yield (TEY) method at different incidence angles $\alpha$ and photon polarizations CL (left circular) and CR (right circular) at $T = 25$ K. We extract the absorption coefficient $\mu$ as described in the Supplementary Material. XAS-MCD is maximized when the helicity vector for the photon is parallel to the sample magnetization. As shown in Fig. 2a, $\mu$ shows no difference between CR and CL at normal incidence ($\alpha = 90°$), while the difference increases with decreasing $\alpha$ from normal to grazing incidence. Fig. 2b shows the normalized differences $[\mu(\alpha)-\mu(90°)]/\cos\alpha$ as a function of $\alpha$, showing a collapse onto two curves of equal magnitude and opposite sign determined by the sign of the photon helicity. Thus, the difference simply follows the factor $\cos\alpha$, consistent with the simplest theory for the variation of XAS-MCD with the angle between the helicity of the photon and sample magnetization $\mathbf{M}$. Using the $\cos\alpha$ factor, we can extrapolate the XAS-MCD to $\alpha = 0°$ (helicity of photon fully parallel (CR) or anti-parallel (CL) to $\mathbf{M}$). Fig. 2c displays the result together with the previous XAS-MCD result for pure iron (thick blue lines)[32]. $Fe_3Sn_2$ shows a larger dichroism at the $L_3$-edge than pure iron. By the sum rules for XAS-MCD, the orbital and spin magnetic moments can be determined[33,34]. We assess the orbital moment *per hole* of an iron site to be $m_{orb} \sim 0.13\,\mu_B$, and the spin moment *per hole* $m_{spin} \cdot (1 + \frac{7\langle T_z \rangle}{2\langle S_z \rangle}) \sim 0.6\,\mu_B$, where $\langle T_z \rangle$ is the expectation value of the magnetic dipole operator and $\langle S_z \rangle$ is equal to half of $m_{spin}$ in Hartree atomic units. The assessed values are for the moment per hole at one Fe atom, so they need to be multiplied by the number of d-orbital holes in an iron atom to obtain the moments per Fe atom. While the spin moment $m_{spin}$ in $Fe_3Sn_2$ is close to the value for pure iron, the orbital moment is much larger, with $m_{orb}/m_{spin} \sim 0.22$ in contrast to 0.043 for pure iron[32].

Using a XAS-MCD sample scan at the absorption maximum, we characterized the moment alignment across the whole sample with (Fig. 2e) and without (Fig. 2f) an in-plane magnetic field. The measurement is done at $\alpha = 20°$. As can be seen, the magnetic field of ~0.13 T polarizes the moments of the whole sample, which shows a large domain with homogeneous XAS-MCD signals. On the other hand, the sample without a magnetic field only shows weak XAS-MCD signals in small and discrete regions due to averaging of the MCD signal over oppositely magnetized domains within the footprint (~ 100 $\mu$m × 5 $\mu$m) of the beam, which is large compared to the magnetic domain size[9].

The large and homogeneous ferromagnetic domain achieved with the external magnetic field also allows measurement of RIXS-MCD for a single magnetic domain. Fig. 2d shows a comparison of the RIXS spectra for samples with and without magnetic field. The former show much more pronounced dichroism than the spectra taken without a field, with a peak around 0.15 eV almost fully suppressed for CL incident polarization. We note that the peak close to zero energy also shows a pronounced dichroism. As non-Bragg elastic scattering has no MCD in a fully polarized crystalline ferromagnet, this peak must be derived from low-energy excitations with origin similar to that of the peak at 0.15 eV, which we identify as magnons in the analysis below.

Figure 3 presents the RIXS results for a single in-plane polarized magnetic domain of $Fe_3Sn_2$ at T = 25 K, which are measured at 130° fixed scattering angle with a $\mathbf{q}$ trajectory shown as the dark-red arc in Fig. 1b and the incident energy tuned to the Fe $L_3$-edge resonance ~ 707 eV. The data were collected at the ADRESS beamline of Swiss Light Source (SLS)[35,36], with instrumental resolution ~ 74 meV full-width-at-half-maximum (FWHM). The displayed spectra are corrected by the self-absorption factors with the outgoing absorption coefficients averaged among different polarizations (see section II of the Supplementary Material). Fig. 3a shows the momentum-dependent spectra for $\mathbf{q} = (H, 0, L(H))$ in a full energy transfer range, while Fig. 3b-d show the spectra for low-energy transfer $\mathbf{q} = (H, 0, L(H))$ and $(H, H, L'(H))$, and with azimuthal ($\phi$) rotation, respectively. The RIXS spectra can be separated into two parts, a broad high-energy peak above 0.4 eV (centered around 2 eV) and low-energy peaks below 0.3 eV. To clarify the nature of these two response components, we measured an incident photon energy dependent RIXS map at $\mathbf{q} = (0.123, 0, 1.97)$, as shown in Fig. 3e. The high-energy peak (~1 eV and larger) shifts to higher energy transfer as the incident energy $E_i$ increases, which is characteristic for 'fluorescence-like'[37] behaviour, while the low-energy peaks stay at fixed energy transfer. The latter Raman-like behaviour suggests a collective nature of the low-energy excitations[38]. A clear RIXS-MCD effect, *i.e.*, different intensities for CL and CR helicity of the incident X-rays, appears for both the low- and high-energy excitations. However, the momentum or incident angle dependences of the MCD for the fluorescence and Raman-responses are markedly different. Fig. 3f displays the integrated MCD intensities in the low- (-0.1 – 0.3 eV) and high- (1.3 – 4 eV) energy ranges of spectra along ($H$, 0) direction (Fig. 3a), which are shown *vs.* the incident angle $\alpha$. The high-energy dichroism follows a $\cos\alpha$ form as does the XAS-MCD, suggesting that it originates mainly from the absorption step in the RIXS process and is trivially proportional to the amount of core holes of the intermediate states created in the absorption step, while the final states are mostly irrelevant. In contrast, the low-energy dichroism depends only weakly on $\alpha$, and still shows a large dichroism even close to normal incidence, where the dichroism in XAS disappears. This suggests that the excited final states are also selectively chosen by the different photon helicities and are responsible for the dichroism together with the initial states. When the sample and magnetic field (fixed along $(H, 0, 0)$ to maintain a fixed magnetic state) are rotated to lie perpendicular to the scattering plane, the entire MCD effect fades out gradually according to $\cos\phi$, as shown in Fig. 3d.

The non-trivial RIXS MCD for the low-energy excitations encodes the specific nature of the final states. In the approximation where core-hole excitations on different sites are uncorrelated, the RIXS cross-section can be written[39] as a sum of local terms representing the resonances multiplied by photon polarization factors and correlation functions formed between magnetization operators at the magnetic atom sites. For spin wave scattering, to lowest order in the magnetization operators, this results in a (zero-temperature) cross-section proportional to

$$I \propto \sum_{\lambda'} \left| \langle \lambda' | \boldsymbol{\varepsilon}_o^* \times \boldsymbol{\varepsilon}_i \cdot \hat{\mathbf{M}}_\mathbf{q} | \lambda \rangle \right|^2 \cdot \delta(E_\lambda - E_{\lambda'} - \hbar\omega) \tag{2}$$

Here, $\boldsymbol{\varepsilon}_i$ and $\boldsymbol{\varepsilon}_o$ are polarization vectors for incident and outgoing photons, respectively. The operator $\hat{\mathbf{M}}_\mathbf{q}$ is the Fourier transform of the local magnetization operator $\hat{\mathbf{M}}(\mathbf{r})$, where $\mathbf{q} = \mathbf{k}_i - \mathbf{k}_o$ is the change of the photon momentum, $|\lambda\rangle$ is the ferromagnetic ground state, and $|\lambda'\rangle$ is an excited state, which can be a magnon. The validity of expression (2) for spin waves observed by RIXS has been extensively tested[39–44]. For RIXS MCD in a Heisenberg ferromagnet, where we sum over outgoing polarizations while taking the difference between the two circular polarizations of the incident X-rays, we use the notation of Fig. 1b for a single domain sample, and obtain (see the details in Methods section):

$$I_{\mathrm{RIXS-MCD}} \propto \sin\beta \cdot \sin(\alpha + \beta) \cdot \cos\phi \cdot Im(S^{zy}(\mathbf{q}, \omega)) \tag{3}$$

where $\beta$ is the angle between incident and outgoing photons, $\alpha$ and $\phi$ are the incident and azimuthal angle, respectively. $S^{zy}(\mathbf{q}, \omega)$ is the $zy$ element of the dynamic spin-spin correlation function, which is imaginary. This is in contrast to neutron scattering, where the off-diagonal elements of $S(\mathbf{q}, \omega)$ cancel each other due to the dipolar polarization factors[45]. For energy loss spectra, i.e., when photons lose energy in the sample to create excitations, $Im(S^{zy}(\mathbf{q}, \omega))$ is equal to $S^{yy}(\mathbf{q}, \omega)$ and $S^{zz}(\mathbf{q}, \omega)$. Equation (3) results in the simple angular dependence, $\sin(\alpha + \beta) \cdot \cos\phi$, for the RIXS MCD of spin wave excitations of a Heisenberg ferromagnet. In Figure 3f, where $\beta$ is fixed at $50°$, we show that the angle dependence of the low-energy excitations follows the $\sin(\alpha + \beta)$ curve very well (see the Supplementary Figure S3 for the angular dependence of the two peaks separately), which confirms the transverse (to the magnetization) spin-wave nature of the excitations.

**Analysis and discussions of the spin waves**

The intensity differences between CR and CL polarizations exclude phonon and elastic scattering, isolating the magnetic contributions to the RIXS cross-section for $Fe_3Sn_2$. By correcting the intensity difference with the polarization factors of spin waves and the self-absorption effect (see section V of the Supplementary Material), we obtain the pure magnetic Raman signals displayed in Fig 4a as a function of in-plane momentum $\mathbf{q}_{//}$. There is a peak with weak momentum dependence around 0.15 eV and a low-energy acoustic mode, which moves away from zero energy as $\mathbf{q}_{//}$ increases from 0 in the quadratic manner expected for ordinary spin waves with a strong exchange interaction within the kagome planes.

To understand the contribution of inter-layer couplings, we measured the out-of-plane momentum ($L$) dependence of the RIXS MCD spectra. Fig. 4c shows the $L$-dependent low-energy RIXS MCD at a fixed in-plane $\mathbf{q}_{//} = (0.04, 0.04)$, measured at the ID32 beamline of European Synchrotron Radiation Facility (ESRF), with a resolution ~ 35 meV FWHM. It clearly illustrates the out-of-phase $L$ dependence of the two modes, exactly as expected for the even and odd modes induced by the bilayer interaction. The even and odd modes represent the in-phase and out-of-phase movements of the spins in the two planes forming the bilayer, and their intensities have $L$ modulations depending on the thickness of the bilayer $z_{bi}$ with the form of $f^2(\mathbf{q}) \cdot \cos^2(\pi z_{bi} L)$ and $f^2(\mathbf{q}) \cdot \sin^2(\pi z_{bi} L)$, respectively, where $f(\mathbf{q})$ is the Fe form factor[28,29], as shown in Fig. 1g. In Fig. 4d, we compare the integrated intensities of the observed two peaks to these simple $L$ modulations (solid lines). The circles are from the ESRF data shown in Fig. 4c at $\mathbf{q}_{//} = (0.04, 0.04)$ integrated over an interval of [-0.07, 0.07] eV (green circles) and [0.07, 0.3] eV (magenta circles), while the plus and cross symbols are from the measurements at the ADRESS beamline of the SLS at $\mathbf{q}_{//} = (0123, 0)$ and $\mathbf{q}_{//} = (0.2, 0)$ shown in Supplementary Figure S5. The observed intensities follow the $L$ modulations very well, except one AC point (green circle) at $L = 1.1$, which drops off the trend due to the degradation of the sample surface at extended time after cleaving, weakening the dichroism of the acoustic mode (see Methods and section VI of the Supplementary Material for the detailed RIXS spectra).

To determine the magnetic interactions and the character of damping, we fit the RIXS MCD spectra shown in Fig. 4a to a convolution of a 74 meV FWHM resolution function with the sum of damped harmonic oscillator profiles:

$$S(\omega) = (n(\omega) + 1) \cdot \frac{A\gamma\omega}{\left(\omega^2 - \omega_0^2\right)^2 + 4\gamma^2\omega^2} \tag{4}$$

where eigenfrequencies are calculated using the spinW package[27] applied to the bilayer Heisenberg model of equation (1). Here, the fitting parameters are $J_1$, $J_{bi}$, the amplitudes, and the damping factors for the relevant spin wave modes (AC, OP1, OP2, and OP3 in Fig. 1d), which show non-zero intensities in the accessed momentum space. The values of $J_1$ and $J_{bi}$ determine the dispersions, which we set to be $\omega_0$ in the fitting. As the OP2 and OP3 modes have negligible intensities at small $\mathbf{q}_\parallel$ (Fig 1d), we manually set them to zero for $\mathbf{q}_\parallel \leq (0.2, 0)$ along $(H, 0)$ and $\mathbf{q}_\parallel \leq (0.12, 0.12)$ along $(H, H)$ to improve convergence. The fitting applies to all the spectra shown in the intensity map of Fig. 4a and gives best results for $J_1 = -25.0 \ (\pm 2.9)$ meV and $J_{bi} = -37.5 \ (\pm 1.8)$ meV. $J_{bi}$ is mostly determined by the minimal energy of the optical OP1 mode, $i.e.$, the splitting energy between the even and odd mode, while $J_1$ is mostly determined by the stiffness of the acoustic mode. The errors are estimated by assuming that the error in elastic peak positions (zero energy transfer) is ~ 10% of the FWHM of the resolution (~7 meV). We plot in Fig. 4b, the simulated spin-spin correlations $Im(S^{zy}(\boldsymbol{q}, \omega) - S^{yz}(\boldsymbol{q}, \omega))$ for a Heisenberg model with the obtained $J_1$ and $J_{bi}$. The $\mathbf{q}$ trajectory is the same as in the experiment and the spectra are broadened by a Gaussian with FWHM = 74 meV. The results look very similar to the data in Fig. 4a. We note that the peak intensity of the acoustic mode seems not to grow along the q trajectory in the experiments (Fig. 4a) as it does in the simulation (Fig. 4b). This apparent discrepancy is a result of the increased damping factor of the acoustic mode in the experiments for larger q values, which will be explained by the fitting results in Fig. 5.

Fig. 5a shows the fitted spectra at four relatively large $\mathbf{q}_\parallel$ along $(H, 0)$, where the hardening of the acoustic mode and emergence of the OP2 mode, which gradually overrides that of the OP1 mode with increasing $\mathbf{q}_\parallel$, yield an excellent (but almost needless to say, not mathematically unique) account of the entire lineshape; attempts to fit these higher momentum transfer data with only two damped harmonic oscillators were not as successful. Fig. 5b and 5c show the integrated intensities and the damping factors of the fitted spin-wave modes, respectively. The damping factor of the acoustic mode increases greatly at larger $\mathbf{q}_\parallel$ values, which reduces the peak intensity and leads to the differing trends in the intensity maps of Fig. 4a and Fig. 4b noted above. Nevertheless, the integrated intensities follow the calculated amplitudes (solid lines) for the $J_1$-$J_{bi}$ model very well, which confirms that the adopted model is appropriate for our results.

Although $J_1$ and $J_{bi}$ may be the most important interactions, other parameters such as the inter-bilayer interaction $J_c$, and a difference between $J_{1a}$ and $J_{1b}$ can exist and modify the spin waves. Spin-space anisotropy and/or $J_c$ would be necessary for the magnetic ordering at a high Curie temperature 640 K[46,47], since there is no order for a pure 2D Heisenberg magnet. $J_c$ will further introduce dispersion along $L$. However, the $L$ dependence (Fig. 4c and section VI of the Supplementary Material) does not reveal dispersion, which suggests $J_c$ is much smaller than $J_1$ and $J_{bi}$. Based both on our measurements concerning the out-of-plane momentum dependence as well as the knowledge of the small anisotropy fields[9], the mean field formula should provide an upper bound[46], which based on considerations of the extended fluctuation regime in 2D systems ending in a quasi-Kosterlitz-Thouless transition can actually be reduced by approximately a factor of two[48]. The outcome of all of our $(H, K, L)$-dependent measurements is $T_{CMF} = \frac{S(S+1)}{3k_B}(z_1 J_1 + z_{bi} J_{bi} + z_c J_c) \sim 1360$ K, implying, when we set $J_c = 0$, an estimated $T_C$ of 680 K, which is remarkably close to the 640 K observed experimentally. Different $J_{1a}$ and $J_{1b}$ exert much subtler influence, mostly on the flatness of the optical modes and their crossing points. The lattice is such that the bonds between Fe atoms do not see an inversion symmetric environment, implying that the Dzyaloshinskii–Moriya interaction (DMI) can be non-zero, as also suggested by the substantial orbital angular momentum associated with the iron atoms. The DMI can open gaps at the crossing points (Fig. 1f), which leads to the question of whether the magnons will become topologically non-trivial. The answer is that the magnon flat band must be topological in the limit of single layers[49] (*i.e.*, where the inter-layer coupling is weak), as it also is for Cu(1,3-bdc)[20]. In the opposite limit where the inter-layer coupling is very strong, we have a triangular lattice with flat modes characteristic of excitations within the octahedral Fe "molecules", and we expect trivial topology. There should therefore be a phase transition between the kagome and triangular lattices as a function of the ratio between the two couplings. The associated effects are beyond the current resolution and momentum transfer range but will be interesting to verify with further RIXS and neutron scattering studies.

Another outcome of the fitting is the exceptionally high spin wave damping shown in Fig. 5c when compared to "local moment" ferromagnets including classic insulators such as EuO[50] and the manganites[51]. In Heisenberg ferromagnets, the spin wave damping can occur due to a diffusion-like process which varies in proportion to $q^4$ for the acoustic mode[52,53]. The solid line in Fig. 5c displays the fitting of the damping of the acoustic mode by $P|\mathbf{q}_{//}|^\delta$, where $|\mathbf{q}_{//}|$ is expressed in dimensionless reciprocal lattice units. The fitted result $\delta = 2.26 \pm 0.60$ is smaller than 4, while $P = 0.40 \pm 0.30$ eV is much larger than not only the small thermal energy $k_B T = 0.002$ eV, but also the observed spin wave energy of the OP1 mode. The damping of the OP1 mode is almost momentum independent with $\gamma \approx 0.027$ eV, measured at SLS, corresponding to a FWHM of 0.055 eV. To account for the strong damping, electronic excitations outside the manifold of spin wave excitations are required. An important origin of the damping could be decay into electron-hole pairs, similar to the high damping in iron and nickel following from spin-conserving (*i.e.*, without appeal to spin orbit effects) decays into electron-hole pairs[54–57]. In Fe₃Sn₂, the electronic states and the Fermi surface respond strongly to the magnetic field and the magnetization direction[13,14,17]. Therefore, the electronic states can be distorted as the spin waves propagate, which in turn damps the spin waves[58]. Furthermore, special regions of the Fermi surface, such as those near band crossings, are lifted by the SOC and depend strongly on the magnetization direction[58]. Indeed, multiple Weyl nodes which are switchable by the magnetization are suggested to exist close to the Fermi level[14]. Spin waves can therefore redistribute Weyl fermions thus shortening the lifetime of both. Despite the strong damping, the high-energy optical spin waves in Fe₃Sn₂ are still well-defined, in contrast to other related kagome magnetic metals such as FeSn[59,60], CoSn[59], and Co₃Sn₂S₂[61–63], in which the high-energy spin waves are much less visible if at all.

**Conclusions**

We have exploited modern synchrotron-based X-ray technology to examine the magnetic order and excitations in the much-celebrated metallic kagome ferromagnet $Fe_3Sn_2$, allowing direct comparison to both insulating kagome ferromagnets and metallic iron. The magnetic circular dichroism of the X-ray absorption reveals that the orbital contribution to the magnetic moment is five times larger than in elemental iron where it is understood to be almost entirely quenched on account of the crystal field energies being larger than the spin-orbit interaction. This is a quantitative manifestation of the large spin-orbit coupling which also makes $Fe_3Sn_2$ a topological material, with numerous Weyl nodes[14], and indeed suggests a method to calibrate the spin-orbit coupling strength SOC introduced "by hand" into DFT: SOC is simply varied to obtain the measured ratio of orbital to spin contributions to the magnetization.

Furthermore, taking advantage of the magnetic circular dichroism of RIXS, we discovered two spin wave bands which are ascribed to the even and odd modes, derived from a strong bilayer coupling, by measurements of the out-of-plane momentum dependence. This means that the underlying magnetic and concomitant electronic Hamiltonians for $Fe_3Sn_2$ are remote from the limit of weakly coupled single kagome layers, thus accounting for the difficulty of finding in both computation (DFT) and experiment (angle resolved photoelectron emission spectroscopy) the flat bands and resolved Dirac points associated with single kagome layers. Another picture emerges from these results, namely that the fundamental low-energy electronic building blocks are triangular lattices of octahedral Fe "molecules", without the possibility for perfectly flat modes in the planar reciprocal space but with many new touching points between the greater number of modes introduced by the "molecules". Our work thus motivates the control of these touching points by environmental and chemical parameters, as well as theory of their topological nature. Finally, that there is strong mixing of the optical modes with the electron-hole pair continuum is clear from their considerable damping even for $q \longrightarrow 0$. The mixing may be due to attempted rearrangements of the $Fe_3Sn_2$ Weyl nodes due to transient magnetization rotations associated with the spin waves. We look forward to more experiments and theory on what happens when topological electrons mix with topological bosons.

## Methods
## Experiments
The XAS, RIXS experiments with 130° fixed scattering angle, and $L$ dependent RIXS measurements at $\mathbf{q} = (0.123, 0, L)$ and $\mathbf{q} = (0.2, 0, L)$ were carried out at the ADRESS beamline of the Swiss Light Source (SLS) at the Paul Scherrer Institut[35,36], while the $L$ dependent RIXS measurements at $\mathbf{q} = (0.04, 0.04, L)$ were done at the ID32 beamline of European Synchrotron Radiation Facility (ESRF). The momentum transfer $\mathbf{q}$ is denoted in reciprocal lattice units (r. l. u.), with lattice constants a = b = 5.315 Å and c = 19.703 Å. For the measurements at ADRESS of SLS, the plate-shaped crystal was cleaved in ultra-high vacuum (~2×10⁻¹⁰ mbar) at $T = 25$ K to yield a clean and flat surface parallel to the $a$-$b$ plane. The scattering plane, spanned by the incident ($\mathbf{k}_i$) and emitted ($\mathbf{k}_o$) photon wave vectors, is perpendicular to the sample $a$-$b$ plane, and intersecting with it for $\mathbf{q}=(H, 0, 0)$ when $\phi = 0$, as shown in Fig. 1b. The instrumental resolution of the RIXS measurements at the ADRESS beamline is ~ 74 meV full-width-at-half-maximum (FWHM) for 130° fixed scattering angle measurements, while it is ~ 80 meV FWHM for $L$ dependent RIXS measurements at 110° and 90° scattering angles. A magnetic field ($\mathbf{B}$ ~ 0.13 T) along the $(H, 0)$ direction from a pair of permanent magnets polarized the ferromagnetic sample. At ID32 of ESRF, the crystal was cleaved in the transfer chamber (~1×10⁻⁸ mbar) at room temperature and transferred immediately into the measuring chamber and cooled down to 25 K with the vacuum ~1×10⁻⁹ mbar; the instrumental resolution is ~ 35 meV FWHM and the magnetic field (~ 0.2 T) is applied along the $(H, H)$ direction. Right (CR) and left (CL) circular polarized incident photons are employed for the measurements, while the polarizations of the emitted photons in RIXS are not resolved. All data were collected at base temperature $\sim 25$ K.

**Calculation of Polarization Factor**

Letting $\mathbf{P} = \boldsymbol{\varepsilon}_o^* \times \boldsymbol{\varepsilon}_i$, equation (2) can be rewritten as:

$$I \propto \sum_{ab} P_a^* P_b \sum_{\lambda'} \langle \lambda | \hat{\mathbf{M}}_{\mathbf{q},a}^\dagger | \lambda' \rangle \langle \lambda' | \hat{\mathbf{M}}_{\mathbf{q},b} | \lambda \rangle \cdot \delta(E_\lambda - E_{\lambda'} - \hbar\omega) \qquad (5)$$

where $a$ and $b$ stand for $x$, $y$, $z$, which are the indices of the vector elements. The formula is very similar to the magnetic scattering of neutrons[45], with only different polarization factors $P_a^* P_b$. As for neutron scattering, the second summation in the above formula is proportional to the spin-spin correlation function $S^{ab}(\mathbf{q}, \omega)$. In the linear approximation for a local moment ferromagnetic system, given that the spins are polarized in the $x$ direction shown by Figure 1b, only the elements $S^{aa}$, $S^{yz}$, and $S^{zy}$ are non-zero. While $S^{xx}$ contributes to the elastic scattering, $S^{yy}$, $S^{zz}$, $S^{yz}$ and $S^{zy}$ are related to the dynamic part, with $S^{yy} = S^{zz}$ and $S^{yz} = -S^{zy}$. In the scattering geometry of Figure 1b, we can define the polarization vectors $\boldsymbol{\varepsilon}_i$ and $\boldsymbol{\varepsilon}_o$. For example, the CL and CR incident polarizations are $(-i\sin\alpha, 1, -i\cos\alpha)/\sqrt{2}$ and $(i\sin\alpha, 1, i\cos\alpha)/\sqrt{2}$, respectively. By summing over the possible outgoing polarizations $\boldsymbol{\varepsilon}_o$, we obtain the cross-section proportional to:

$$I_{CL/CR} \propto [(\sin\alpha)^2 + (\sin(\alpha+\beta))^2] \cdot S^{zz}(\mathbf{q}, \omega) + (\sin\beta)^2 \cdot S^{yy}(\mathbf{q}, \omega)$$
$$(-/+)\sin\beta \cdot \sin(\alpha+\beta) \cdot iS^{yz}(\mathbf{q}, \omega)(+/-)\sin\beta \cdot \sin(\alpha+\beta) \cdot iS^{zy}(\mathbf{q}, \omega) \qquad (6)$$

For energy loss spectra like RIXS, $S^{zy}$ is imaginary and equal to $iS^{yy}$. If we further include the azimuthal rotation, a $\cos\phi$ factor is needed, which altogether results in the RIXS MCD formula as equation (3).

**Data availability**

The data that support the findings of this study are shown in the main text figures and the supplementary data figures. The source data generated during and/or analyzed in the current study will be deposited in a public repository before publishing (accession codes will be updated before publication).

**Acknowledgements**: The experiments were performed at the ADRESS beamline of the Swiss Light Source at the Paul Scherrer Institut (PSI) and the ID32 beamline of European Synchrotron Radiation Facility (ESRF) (proposal no. HC-5443). The experimental work at PSI is supported by the Swiss National Science Foundation through project no. 178867, 207904 and the Sinergia Network Mott Physics Beyond the Heisenberg Model (MPBH) (SNSF Research Grant 160765). T.C.A., Y.W., and C.W.G. acknowledge funding from the European Union's Horizon 2020 research and innovation programme under the Marie Sklodowska-Curie grant agreement No.701647 and 884104 (PSI-FELLOW-II-3i program). T.Y., Y.S., and T.S. acknowledge the "Cross" funding from Paul Scherrer Institut. G.A. and W.Z. were supported by the European Research Council under the European Union's Horizon 2020 research and innovation programme HERO (Grant agreement No. 810451). Y.X. acknowledges funding from the Innovation Program for Quantum Science and Technology (No. 2021ZD0302802).

**Competing interests**: The authors declare no competing interests.

**Author contributions**: G.A. conceived the project. J.L. grew and characterized the single crystals with support from Y.X. W.Z., T.C.A., Y.T., Y.W., C.W.G., Z.Z., and T.S. performed the XAS and RIXS experiments at Swiss Light Source in discussion with G.A. and Y.S and support from V.S. W.Z., Y.W., T.Y., K.K., and T.S. performed the RIXS experiments at European Synchrotron Radiation Facility. W.Z. analyzed the data in discussion with Y.S., T.S., and G.A. W.Z., Y.S., T.S., and G.A. wrote the manuscript. Y.S., T.S., and G.A. coordinated the research.

**Additional Information**: Supplementary Information is available for this paper.

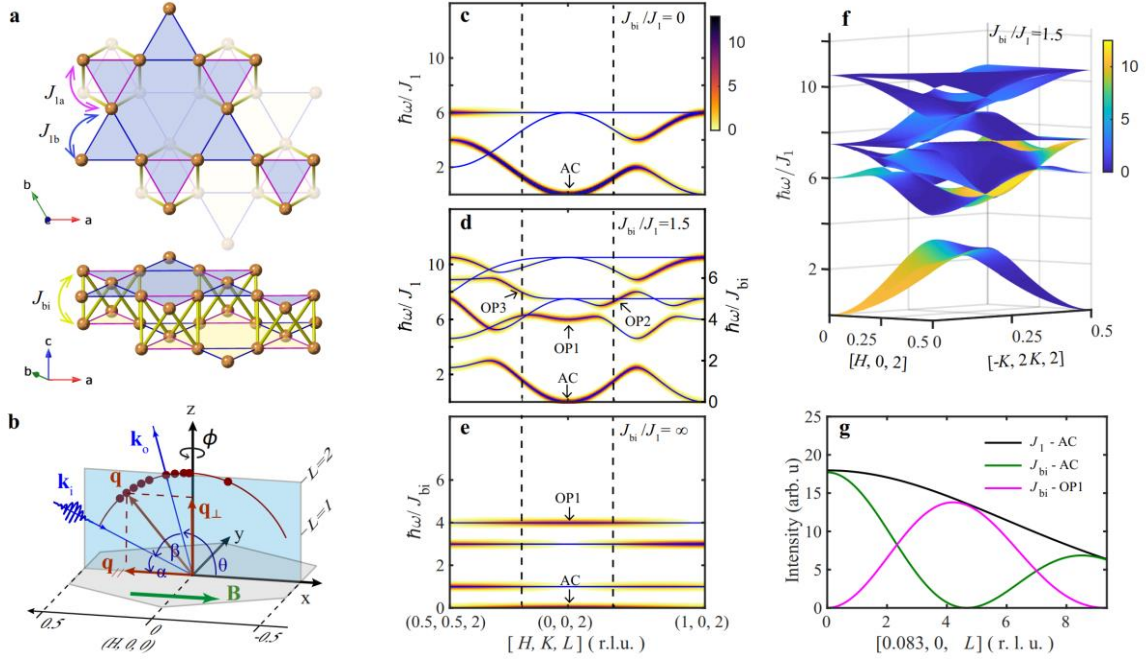

Fig. 1 **Kagome bilayer of Fe$_3$Sn$_2$, experimental geometry, and spin waves for an AB kagome bilayer**. **a**, Top (upper) and side (bottom) view of kagome bilayer structure of Fe atoms in Fe$_3$Sn$_2$. The arrows illustrate the potentially major exchange couplings in the bilayer: the in-plane nearest-neighbour interaction $J_{1a}/J_{1b}$, with a (b) indicating the shorter (longer) bond in the breathing kagome lattice, and the bilayer coupling $J_{bi}$. **b**, Scattering geometry and sample orientation in the experiment. The gray hexagon represents the first Brillouin zone in $L = 0$ plane, where $H$, $K$, $L$ are the Miller indexes in reciprocal lattice units (r. l. u.). $\mathbf{k}_i$ and $\mathbf{k}_o$ are the incident and emitted X-rays, while $\alpha$ and $\theta$ are the incident and emission angle, respectively. $x$, $y$, $z$ are directions of sample translation in sample scans displayed in Fig. 2**e** and **f**. An in-plane magnetic field **B** along the green arrow is applied to the sample. $\mathbf{q}$ is the total momentum transfer, $\mathbf{q}_{//}$ and $\mathbf{q}_{\perp}$ are the projections of $\mathbf{q}$ in the sample $H$-$K$ plane and $(0, 0, L)$ direction, respectively. The arc indicates the momentum trajectory in reciprocal space for a fixed scattering angle 180°- $\beta$ = 130° (dark red), and the dots indicate the measured momentum points in the ($H$, 0, $L$) plane. **c**-**e** show the simulated spin wave dispersions and spin-spin correlations (in momentum space) based on linear spin wave theory for different $J_{bi}/J_1$ ratio in $L = 2$ plane; **c** and **e** represent the single plane and single octahedral Fe molecule limits, respectively, while **d** is for the multiple couplings established in the present experiment. The blue solid lines are the dispersions, and the color maps indicate the strength of the spin-spin correlations. The dashed lines indicate the boundaries of in-plane momentum that can be reached for 130° fixed scattering angle. **f**, the 2D spin wave dispersion and spin-spin correlations in [$H$, $K$, 2] plane for $J_{bi}/J_1 = 1.5$. **g**, $L$ dependence of the intensities of spin wave modes at fixed $\mathbf{q}_{//} = (0.083, 0)$: the dark line indicates the acoustic (AC) mode with only $J_1$ interaction, while the green and magenta lines indicate the acoustic (AC) mode and the bilayer split optical mode (OP1) with $J_{bi}$ interaction.

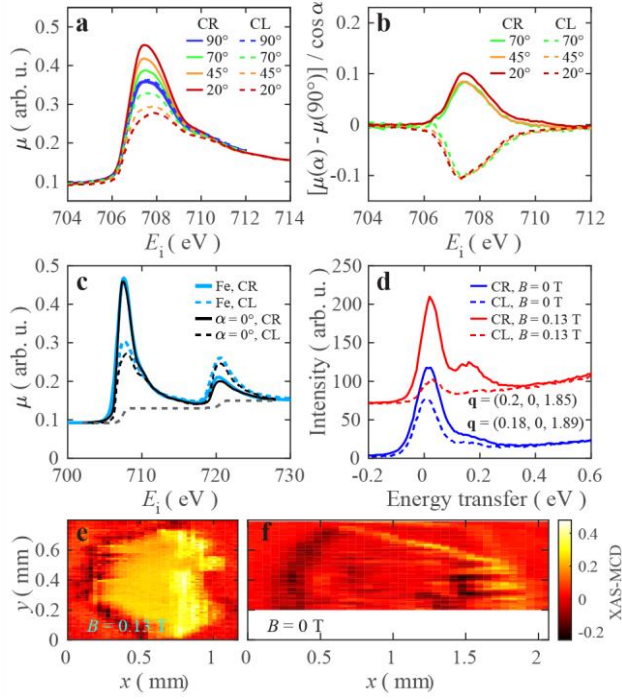

Fig. 2 **Magnetic circular dichroism of X-ray absorption spectra. a**, Absorption coefficients $\mu$ of $Fe_3Sn_2$ under an in-plane magnetic field ($\sim 0.13$ T) at different incident angles ($\alpha$) and circular polarizations (CR and CL), extracted from X-ray absorption spectra measured by total electron yield (TEY) method at $T = 25$ K (see Supplementary Material). **b**, The difference between $\mu$ at a certain angle $\alpha$ and $\mu$ at $\alpha = 90°$, scaled by a factor of $\cos\alpha$. **c**, The extrapolated absorption coefficients $\mu$ ($\alpha = 0°$) of $Fe_3Sn_2$ (black lines) for CR and CL polarization. The thick blue lines are the absorption coefficients of pure iron[32]. The gray dashed line is a two-step-like function for removing $L_3$ and $L_2$ edge jumps. The height of the step at $L_3$ is twice the height at the $L_2$ edge[32]. **d**, RIXS spectra measured at Fe $L_3$ resonance and $T = 25$ K for a sample with in-plane magnetic field (red curves, $\alpha = 40°$) and without magnetic field (blue curves, $\alpha = 43°$). **e** and **f** show the XAS-MCD signals (($TEY_{CR} - TEY_{CL}$) / ($TEY_{CR} + TEY_{CL}$)) at Fe $L_3$-edge resonance scanned across the sample surface with and without magnetic field, respectively. The incident angle $\alpha$ is 20°. The scan step along x is 0.03 mm and 0.06 mm for **e** and **f**, respectively, and 0.01 mm along y for both. The x and y directions are depicted in Fig. 1**b**.

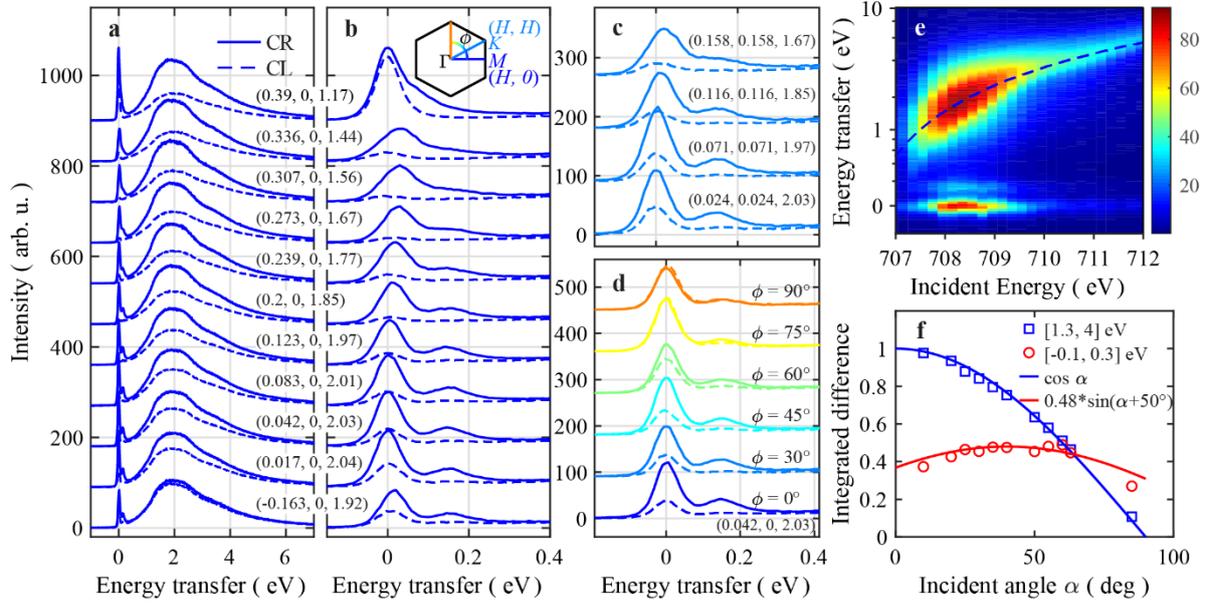

Fig. 3 **RIXS spectra with momentum and incident energy dependence. a-d**, RIXS spectra with CR and CL polarizations at Fe $L_3$ resonance and $T = 25$ K: **a** and **b** along $(H, 0, L)$ direction, **c** along $(H, H, L)$, and **d** at different azimuthal angles. Here the scattering angle is fixed to 130°, so the out-of-plane momentum $\mathbf{q}_\perp$ varies as shown by the dark red arc in Fig 1**b**. The inset hexagon in **b** indicates the Brillouin zone and the lines indicate the in-plane momentum directions. **e**, The RIXS intensity map as a function of energy transfer and incident energy for CR polarization; the y-axis is logarithmic, and the dashed line indicates a linear dependence on the incident energy. **f**, Integrated intensities of RIXS MCD in **a**, in an energy interval [1.3 eV, 4 eV] (blue squares) and [-0.1 eV, 0.3 eV] (red circles) as a function of incident angle $\alpha$. The blue line is a curve of $\cos\alpha$, and the red line is $0.48 \cdot \sin(\alpha + 50°)$. The integrated intensities are divided by the size of the integrated energy interval and normalized by a scaling factor so that the fitting prefactor of the blue squares to $\cos\alpha$ is 1.

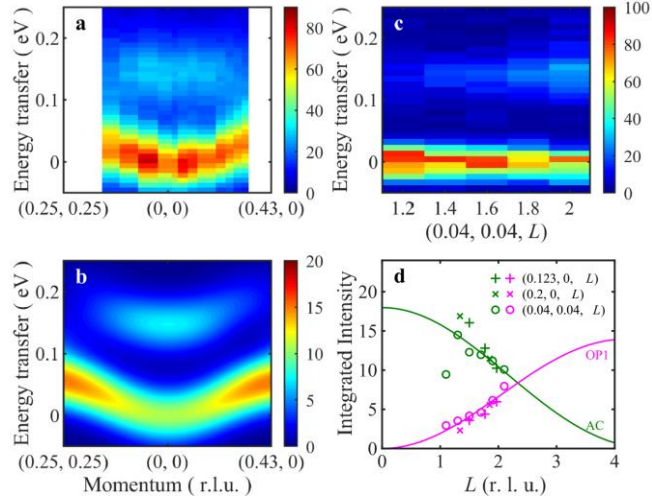

Fig. 4 **Spin excitations in q–E space compared to Heisenberg $J_1$-$J_{bi}$ model. a**, intensity map of RIXS intensity differences between CR and CL polarizations (RIXS MCD) as a function of momentum; the data are from the fixed-scattering-angle (130°) measurements as shown in Fig. 3. The results are already corrected by polarization and self-absorption factors as described in the Supplementary Material. The out-of-plane momentum $L$ changes as $q_{//}$ changes, as shown by the red points on the red arc in Fig. 1b. **b**, Spin-spin correlations $Im(S^{zy}(\boldsymbol{q}, \omega) - S^{yz}(\boldsymbol{q}, \omega))$ simulated for a Heisenberg model with $J_1$ = -25 meV and $J_{bi}$ = -37.5 meV with the same $\boldsymbol{q}$ trajectory and resolution broadening (74 meV) as in **a**. **c**, the out-of-plane $L$ dependence of the RIXS MCD at fixed $q_{//}$ = (0.04, 0.04), measured at ESRF with instrument resolution ~ 35 meV at Fe $L_3$ resonance and $T$ = 25 K. **d**, Integrated intensities of the low-energy acoustic modes (AC, green symbols) and the optical modes (OP1, magenta symbols) as a function of out-of-plane momentum $L$. The solid lines are the expected $L$ modulations for the even and odd modes arising from the inter-plane couplings in the bilayers (bilayer interaction).

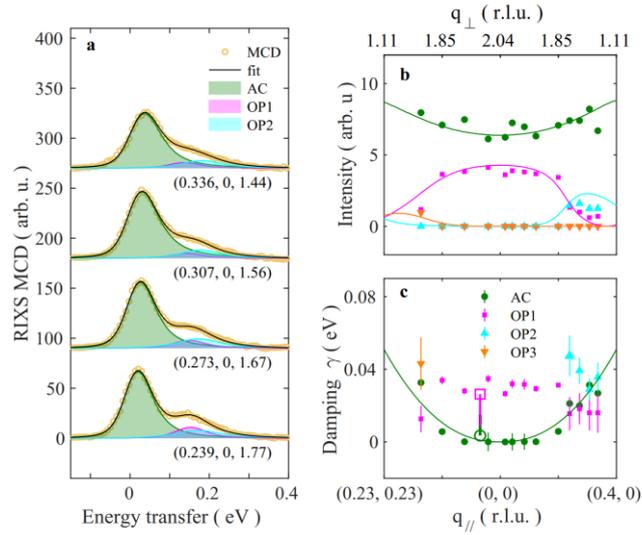

Fig. 5 **Spin-wave theory description of the RIXS magnetic circular dichroism. a**, examples of fitting by $J_1$-$J_{bi}$ model with damped harmonic oscillator profiles on the RIXS MCD at four relatively large $\mathbf{q}_\parallel$ along ($H$, 0). The dispersion derives from $J_1$ and $J_{bi}$ being fixed as described in the text while the amplitude and damping are allowed to vary in the fits. The complete fitting for all momenta is presented in Supplementary Figure S6. **b**, the integrated intensities of the fitted spin-wave modes; the green circles and magenta squares indicate the even acoustic mode (AC) and odd optical mode (OP1), respectively. The cyan and orange triangles indicate the other optical modes that emerge at relatively large $\mathbf{q}_\parallel$ along ($H$, 0) and ($H$, $H$), respectively, as shown in Fig. 1**d**. The solid lines indicate the simulated intensities of the spin wave modes based on the $J_1$-$J_{bi}$ model. **c**, the fitted damping factors for the four spin wave modes. The thin bars represent the standard errors from the fitting procedure. The green solid line corresponds to P$|\mathbf{q}_\parallel|^\delta$ with $\delta$ = 2.26, P = 0.40 eV, and $|\mathbf{q}_\parallel|$ in reciprocal lattice units defined such that (1, 0) occurs at 1.365 Å$^{-1}$. The open circle and square are the fitted damping factors of AC and OP1 modes of $L$ dependent data from ESRF at $\mathbf{q}$ = (0.04, 0.04, 2.1), respectively. The thick bars indicate the range of fitted damping factors for the $L$ dependent data at $\mathbf{q}$ = (0.04, 0.04, $L$) with $L$ varying from 2.1 to 1.1, detailed values are shown in Supplementary Figure S4e.